\documentclass[reprint,pra,superscriptaddress,twocolumn,showpacs]{revtex4-1}
\usepackage{graphicx,color}
\usepackage{amsmath,amssymb,bm}
\usepackage{braket}		

\usepackage[plainpages=false,pdfpagelabels,colorlinks=true,linkcolor=red,urlcolor=blue,citecolor=blue,pdftitle={Title},pdfauthor={},pdfdisplaydoctitle=true,pdfduplex=DuplexFlipLongEdge]{hyperref}

\usepackage[normalem]{ulem}

\definecolor{darkred}{rgb}{0.90,0.2,0.2}
\definecolor{darkgreen}{rgb}{0,0.60,.2}
\definecolor{darkblue}{rgb}{0.1,0.3,1}
\definecolor{grey}{cmyk}{0,0,0,0.25}
\definecolor{orange}{cmyk}{0,0.6,0.8,0}

\begin{document}
\title{ Dynamical Quasicondensation of Hard-Core Bosons at Finite Momenta}

\author{L. Vidmar}
\affiliation{Arnold Sommerfeld Center for Theoretical Physics, Ludwig-Maximilians-Universit\"at M\"unchen, D-80333 M\"unchen, Germany}
\affiliation{Department of Physics, Ludwig-Maximilians-Universit\"at M\"unchen, D-80333 M\"unchen, Germany}
\author{J. P. Ronzheimer}
\author{M. Schreiber}
\author{S. Braun}
\author{S.~S.~Hodgman}
\affiliation{Department of Physics, Ludwig-Maximilians-Universit\"at M\"unchen, D-80333 M\"unchen, Germany}
\affiliation{Max-Planck-Institut f\"{u}r Quantenoptik, D-85748 Garching, Germany}
\author{S. Langer}
\affiliation{Department of Physics and Astronomy, University of Pittsburgh, Pittsburgh, Pennsylvania 15213, USA}
\author{F. Heidrich-Meisner}
\affiliation{Arnold Sommerfeld Center for Theoretical Physics, Ludwig-Maximilians-Universit\"at M\"unchen, D-80333 M\"unchen, Germany}
\affiliation{Department of Physics, Ludwig-Maximilians-Universit\"at M\"unchen, D-80333 M\"unchen, Germany}
\author{I. Bloch}
\affiliation{Department of Physics, Ludwig-Maximilians-Universit\"at M\"unchen, D-80333 M\"unchen, Germany}
\affiliation{Max-Planck-Institut f\"{u}r Quantenoptik, D-85748 Garching, Germany}
\author{U. Schneider}
\affiliation{Department of Physics, Ludwig-Maximilians-Universit\"at M\"unchen, D-80333 M\"unchen, Germany}
\affiliation{Max-Planck-Institut f\"{u}r Quantenoptik, D-85748 Garching, Germany}
\affiliation{Cavendish Laboratory, University of Cambridge, J.\ J.\ Thomson Avenue, Cambridge CB3 0HE, United Kingdom}

\begin{abstract}
Long-range order in quantum many-body systems is usually associated with equilibrium situations.
Here, we experimentally investigate the quasicondensation of strongly-interacting bosons at finite momenta in a far-from-equilibrium case.
We prepare an inhomogeneous initial state consisting of one-dimensional Mott insulators in the center of otherwise empty one-dimensional chains in an optical lattice with a lattice constant $d$.
After suddenly quenching the trapping potential to zero, we observe the onset of coherence in spontaneously forming quasicondensates in the lattice.
Remarkably, the emerging phase order differs from the ground-state order and is characterized by peaks at finite momenta $\pm (\pi/2) (\hbar / d)$ in the momentum distribution function.
\end{abstract}

\maketitle


The nonequilibrium dynamics of quantum many-body systems constitutes one of the most challenging and intriguing topics in modern physics.
Generically, interacting many-body systems are expected to relax towards equilibrium and eventually thermalize~\cite{rigol08,polkovnikov11}.
This standard picture, however, does not always apply. 

In open or driven systems, one fascinating counterexample is the emergence of novel steady states with far-from-equilibrium long-range order, i.e., order that is absent in the equilibrium phase diagram.
This includes lasers~\cite{chiocchetta15}, where strong incoherent pumping gives rise to a coherent emission, and driven ultracold atom systems~\cite{vorberg13}.
The emergence of order far from equilibrium is also studied in condensed matter systems~\cite{stojchevska14} and optomechanical systems~\cite{ludwig13}.

In recent years, closed quantum systems without any coupling to an environment have come into the focus of experimental and theoretical research.
Experimental examples range from ultracold atoms~\cite{greiner02,kinoshita06,hofferberth07,trotzky12,cheneau12,schneider12,ronzheimer13,langen13,Xia2015} to quark-gluon plasmas in heavy ion collisions~\cite{berges15}.
In closed, non-driven systems, two famous examples for the absence of thermalization~\cite{kinoshita06,hofferberth07,langen13} are many-body-localized~\cite{nandkishore14,altman14,schreiber15} and integrable systems~\cite{rigol07}.
These peculiar systems allow for nonergodic dynamics and novel quantum phenomena.

Spontaneously emerging order is in general associated with equilibrium states at low temperatures.
The canonical example is the emergence of (quasi-) long-range phase coherence when cooling an ideal Bose gas into a Bose-Einstein (quasi-) condensate~\cite{anderson95,davis95}.
In this case, thermodynamics ensures that, for positive temperatures~\cite{braun13}, the single-particle ground state becomes macroscopically occupied and thereby dictates the emerging order.
Even in studies of the nonequilibrium dynamics at quantum phase transitions~\cite{sachdevbook}, the emergence of coherence is typically associated with gently crossing the transition from an unordered into an ordered  state, and the strongest correlations and largest coherence lengths appear in the adiabatic limit~\cite{chenD11,braun15}.

\begin{figure}[!bt]
\begin{center}
\includegraphics[width=0.95\columnwidth]{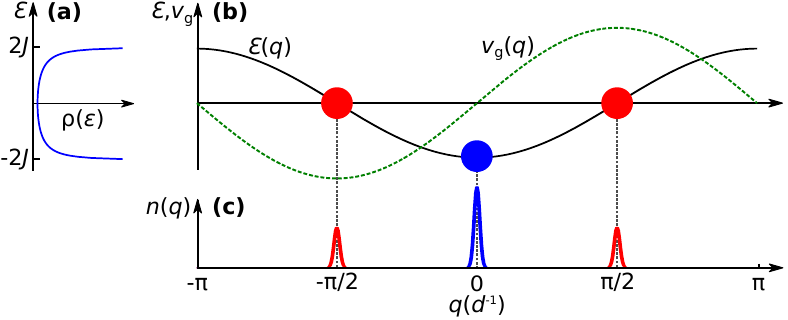}
\caption{
{\it Quasicondensation of bosons.}
(a) Density of states $\rho(\epsilon)$ of a homogeneous 1D lattice.
(b) Dispersion $\varepsilon(q)$ (solid line) and group velocity $v_g(q)$ (dotted line) versus quasimomentum $q$.
(c) Sketch of quasimomentum distribution $n(q)$:
In equilibrium, 1D bosons quasicondense at the minimum of the band at $q=0$, while
in a sudden expansion the quasicondensation of hard-core bosons occurs at $\hbar q=\pm (\pi/2)(\hbar/d)$.
This quasimomentum lies in the middle of the spectrum and is consistent with the vanishing energy per particle of this closed many-body system.
} 
\label{fig1}
\end{center}
\end{figure}

Here, in contrast, we study a condensation phenomenon of strongly interacting lattice bosons far from equilibrium.
After a sudden quantum quench, we experimentally observe the spontaneous emergence of a long-lived phase order that is markedly different from the equilibrium order (cf.\ Fig~\ref{fig1}).
To this end, we prepare a density-one Mott insulator of strongly interacting bosons in the center of a three-dimensional (3D) optical lattice.
Next, we  transform the system into  an array of independent  one-dimensional (1D) chains, entering the regime of integrable hard-core bosons (HCBs).
By suddenly quenching the confining potential along the chains to zero, we induce a sudden expansion of the cloud in a homogeneous lattice~\cite{schneider12,ronzheimer13,reinhard13,Xia2015} with a lattice constant $d$ and detect the formation of a non-ground-state phase profile as a dynamical emergence of peaks at momenta $\hbar k=\pm (\pi/2) (\hbar/d)$, half way between the middle and the edge of the Brillouin zone, in time-of-flight (TOF) distributions.
This finite-momentum quasicondensation was first discussed by Rigol and Muramatsu~\cite{rigol04} (see also Refs.~\cite{daley05,rigol05a,rodriguez06,hm08,vidmar13}), but  has not been studied experimentally so far.

\begin{figure}[!]
\begin{center}
\includegraphics[width=0.99\columnwidth]{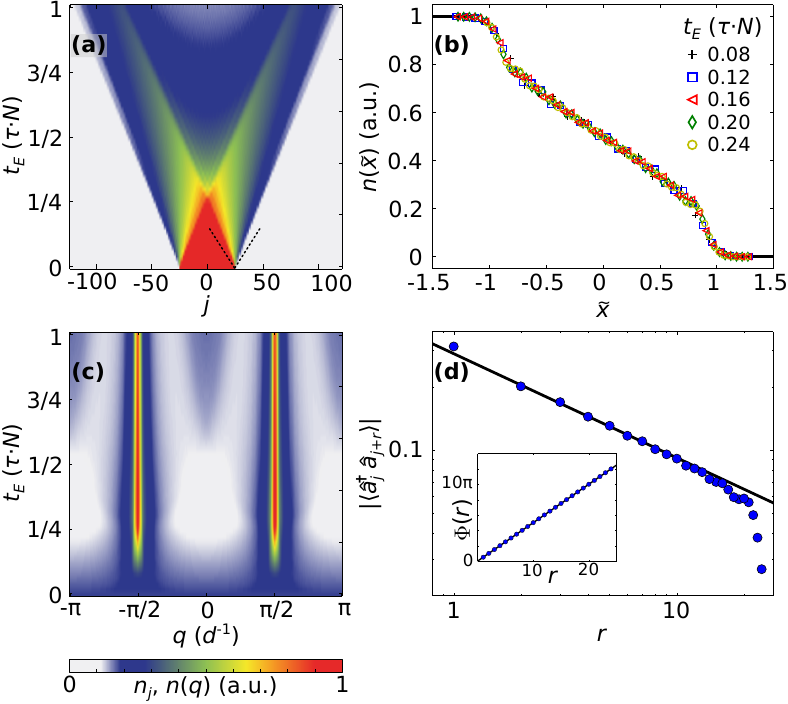}
\caption{
{\it Finite-momentum quasicondensation under ideal conditions}~\cite{rigol04}.
We consider $N=50$ initially localized HCBs.
(a) Density $n_j$ as a function of time.
(b) Density as a function of the rescaled coordinate $\tilde{x}=(j-25.5)/(2t_{\rm E}/\tau)$ in the region bounded by the dashed lines in (a).
For $\tilde{x} \leq 1$, the data collapse to the scaling solution~\cite{antal99} $n(\tilde{x}) = \arccos{(\tilde{x})}/\pi$.
(c) Quasimomentum distribution $n(q)$ as a function of time.
(d) One-particle correlations at $t_{\rm E}=0.24 N \tau$.
Main panel: $|\langle \hat a_j^\dagger \hat a_{j+r} \rangle|$ at $j=26$ (circles) and ${\cal A}(r) = \alpha/\sqrt{r}$ (line) with $\alpha=0.29$~\cite{ovchinikov07}.
Inset: Phase pattern $\Phi(r)$.
} 
\label{fig2}
\end{center}
\end{figure}

{\it Ideal case.}
The idealized setup to study finite-momentum quasicondensates is shown in Fig.~\ref{fig2}.
We consider the Hamiltonian $H=-J\sum_j (\hat a^{\dagger}_{j+1} \hat a_j + {\rm h.c.})$, where $\hat a^{\dagger}_j$ creates a HCB on site $j$ 
of a 1D lattice.
The infinitely large on-site repulsion is accounted for by the hard-core constraint $(\hat a_j^{\dagger})^2=0$.
The initial state is a product state $|\psi_0\rangle = \prod_{j \in L_0} \hat a^{\dagger}_j |\emptyset\rangle $,
completely filling the central $L_0$ sites of an otherwise empty  and infinitely large 1D lattice.
This initial state consists of $N=L_0$ localized particles with a flat quasimomentum distribution and contains no off-diagonal correlations, i.e., $\langle \hat a^\dagger_j \hat a_{j+r}\rangle = 0$ for $r \neq 0$.
Surprisingly,  the quasimomentum distribution
$n(q)=\frac{1}{L} \sum_{j,l} e^{-iq(j-l)d} \langle \hat a_j^\dagger \hat a_l \rangle$
develops singularities at {\it finite} quasimomenta $\hbar q=\pm(\pi/2) (\hbar/d)$ during the expansion  ($t_{\rm E}>0$).
As shown in Fig.~\ref{fig2}(c), these singularities correspond to the emergence of {\it power-law}  correlations
\begin{equation} \label{phaseorder}
\langle \hat a^\dagger_j \hat a_{j+r}\rangle = {\cal{A}}(r) e^{i \Phi(r)};\enspace  {\cal{A}}(r)\sim r^{-\frac{1}{2}}; \enspace \Phi(r) = \pm\frac{\pi}{2} r \;
\end{equation}
in each half of the expanding cloud, shown in Fig.~\ref{fig2}(d).
These power-law correlations justify the name quasicondensate~\cite{rigol04}.
Curiously, the exponent $1/2$ equals the {\it ground-state} exponent~\cite{rigol04,rigol05a}, even though the system is far away from equilibrium, with the energy per particle being much higher than in  the ground state.
In contrast to the ground state, the correlations show a running phase pattern $\Phi(r)$ with a  phase difference of $\pm\pi/2$ between neighboring lattice sites, giving rise to peaks at finite quasimomenta.
Coherence and quasicondensation emerge independently in the left- and right-moving halves of the cloud, corresponding to two macroscopically occupied degenerate eigenstates of the one-particle density matrix $\langle \hat a_j^\dagger \hat a_l \rangle$ that have spatial support in the left- or right-moving cloud, respectively~\cite{rigol04}.
This quasicondensation at finite quasimomenta can equivalently be seen as quasicondensation at $q=0$ in the respective co-moving frames~\cite{hm08}.

In one dimension, HCBs can be exactly mapped to noninteracting spinless fermions via the  Jordan-Wigner transformation~\cite{cazalilla11}.
By virtue of this mapping, the density  $n_j = \langle \hat a_j^\dagger \hat a_j \rangle$ of HCBs is  identical to that of free fermions for all times, whereas the same is not true for the quasimomentum distribution~\cite{paredes04,kinoshita04}.
While the occupations of fermionic quasimomenta are constants of motion, the non-local phase factors in the Jordan-Wigner transformation give rise to the intricate momentum dynamics studied here.
In the ideal case, the dynamical quasicondensates form over a time scale $t_{\rm E}^* \sim 0.3 N \tau$~\cite{rigol04,vidmar13}, where $N$ is the number of particles in the initial state and $\tau = \hbar/J$ denotes the tunneling time.
For very long times, $n(q)$ slowly decays back to its original flat form as a consequence of the dynamical fermionization mechanism \cite{rigol05,minguzzi05,vidmar13}.

\begin{figure*}[!bt]
\begin{center}
\includegraphics[width=1.99\columnwidth]{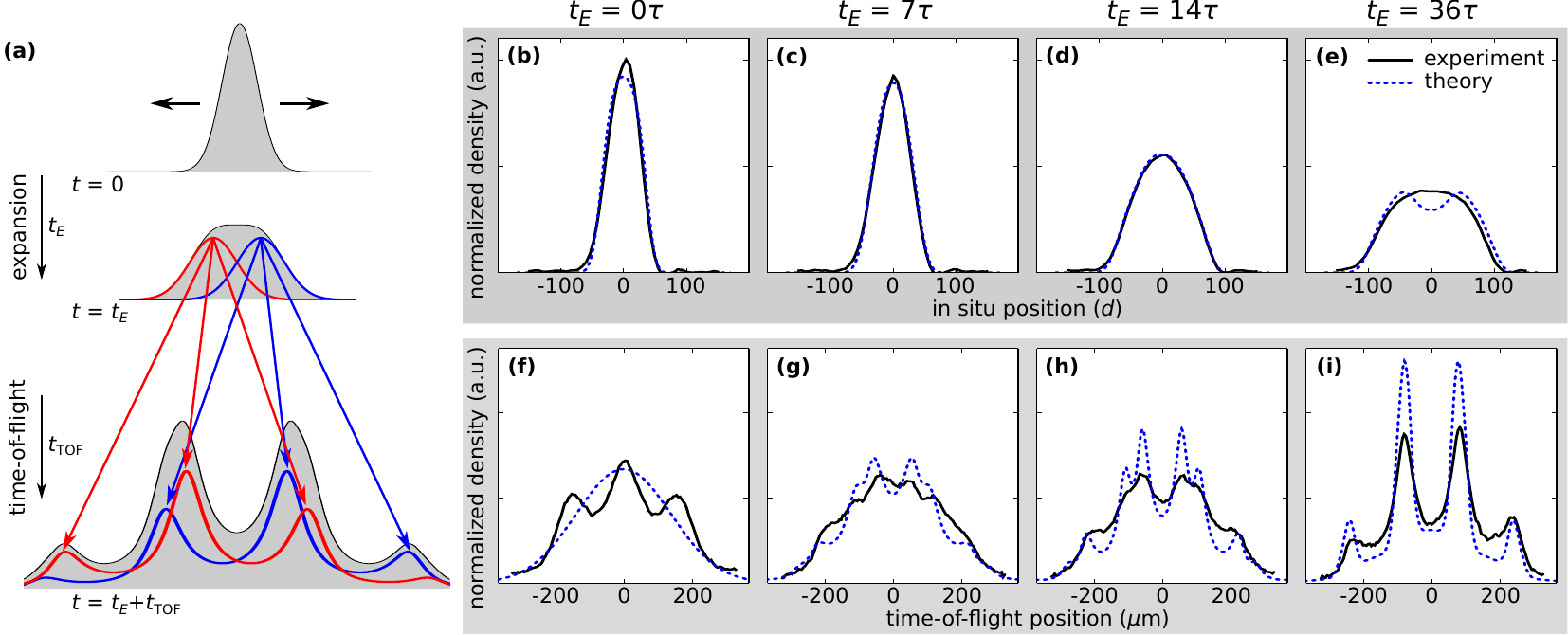}
\caption{
{\it Dynamics during sudden expansion and time-of-flight sequence.}
(a) Sketch of the experimental sequence: We start with a trapped gas that is released from the initial confinement at $t=0$ (top) and then expands for a time $t_{\rm E}$ \textit{in the optical
lattice} (middle). At $t_{\rm E}$, all potentials are removed and the distribution is measured after a finite time-of-flight time $t_{\rm TOF}$ (bottom).
(b)-(e): {\it In-situ} density distributions during the sudden expansion, integrated along the $y$- and $z$-axes.
The lattice constant corresponds to $d=\lambda/2\approx368\,$nm.
(f)-(i): TOF density distributions taken at  $t_{\rm TOF}=6\,$ms for the same $t_{\rm E}$ as in (b)-(e).
All measured density distributions are averaged over 9 to 11 experimental realizations.
} 
\label{fig3}
\end{center}
\end{figure*}

The dynamical quasicondensation at finite momenta is an example of a more general emergence of coherence in a sudden expansion.
For instance, interacting fermions described by the Fermi-Hubbard model exhibit ground-state correlations
in the transient dynamics as well~\cite{hm08}.
Furthermore, there is a close connection to quantum magnetism, as can be seen by mapping HCBs to a spin-1/2 XX chain:
The transient dynamics in each half of the expanding cloud of HCBs is
 equivalent to the melting of a domain-wall state~\cite{antal99,gobert05,platini07,antal08,lancaster10,santos11,eisler13,sabetta13,halimeh14,alba14}  of the form 
$|\psi_0\rangle = $ $|$$\uparrow$$\dots$$\uparrow \uparrow \downarrow \downarrow$$\dots$$\downarrow \rangle$.
For this problem,  a  scaling solution exists~\cite{antal99}, which also applies to the sudden expansion at times $t<t_{\rm E}^*$.
As a consequence, the densities $n_j$ measured at different times collapse onto a single curve, as shown in Fig.~\ref{fig2}(b).
Furthermore, for spin-1/2 XX chains the emergence of power-law decaying transverse spin correlations modulated with a phase of $\pi/2$ has been derived analytically~\cite{lancaster10}.
An interesting perspective onto the emergence of coherence results from  noticing that both our expanding bosons and the melting domain-walls realize current-carrying nonequilibrium steady states (see~\cite{suppmat} for a discussion).

{\it Experimental set-up and results.}
The experimental set-up is identical to that employed in our previous experiment on {\it in-situ} density dynamics~\cite{ronzheimer13}.
We load a Bose-Einstein condensate of approximately 10$^5$ $^{39}\mathrm{K}$ atoms from a crossed optical dipole trap into a blue-detuned 3D optical lattice with a lattice depth of $V_0\approx 20E_r$, where $E_r=h^2/(2m\lambda^2)$ denotes the recoil energy with atomic mass $m$ and lattice laser wavelength $\lambda\approx 737\,$nm.
During the loading of the lattice, we use a magnetic Feshbach resonance to induce strong repulsive interactions between the atoms,  suppressing the formation of doubly occupied sites~\cite{ronzheimer13}.
This results in a large density-one Mott insulator in the center of the cloud.
We hold the atoms in the deep lattice for a $20\,$ms dephasing period, during which residual correlations between lattice sites are mostly lost such that the atoms essentially become localized to individual lattice sites~\cite{Will2010}.
The expansion is initiated at $t_{\rm E}=0$ by simultaneously lowering the lattice depth along the expansion axis in $150\,\mu\mathrm{s}$ to $V_0^{x}\approx 8\, E_r$ (setting $J\approx h \times 300\,$Hz,  $\tau \approx0.5\,$ms), and reducing the strength of the optical dipole trap  to exactly compensate the anti-confinement of the blue-detuned lattice beams.
This creates a flat potential along the expansion direction.
Figures~\ref{fig3}(b)-\ref{fig3}(e) show the ballistic expansion of the {\it in-situ} density~\cite{ronzheimer13}, monitored using absorption imaging.
During the deep lattice period, the magnetic field is changed to tune the on-site interaction strength during the expansion to $U=20J$.
We have numerically verified that the essential features of dynamical quasicondensation are still present for this value of $U/J$, with a shift of the peak position by $\approx10$\% towards smaller values~\cite{rodriguez06,suppmat}.

In order to measure the momentum distribution as a function of expansion time $t_{\rm E}$, we employ an adapted TOF imaging technique.
Directly before shutting off all lattice and trapping potentials, we rapidly increase the lattice intensity along the expansion axis for $5\,\mu\mathrm{s}$ to a depth of $33E_r$.
This time is too short to affect correlations between different sites and the momentum distribution.
Nonetheless, it results in a narrowing of the Wannier functions, which leads to a broadening of the Wannier envelope in the TOF density distribution and thereby facilitates the observation of higher-order peaks~\cite{suppmat}.

Figures~\ref{fig3}(f)-\ref{fig3}(i) contain the main result of our experiment, namely the TOF density distributions, which correspond approximately to the momentum distribution, taken at different expansion times $t_{\rm E}$.
In Fig.~\ref{fig3}(f) the TOF sequence was initiated at $t_{\rm E}=0$, i.e., directly after initiating the expansion.
We observe a central peak at $k=0$ and two higher order peaks at $\hbar k=\pm (2\pi)(\hbar/d)$, indicating a weak residual $k=0$ coherence probably resulting from an imperfect state preparation. 
During the expansion, however, the momentum distribution changes fundamentally and the remnants of the initial coherence quickly vanish.
Instead, new peaks at finite momenta are formed.
These peaks directly signal the spontaneous formation of a different phase order.
This is best seen in Fig.~\ref{fig3}(i) at $t_{\rm E}=36 \tau$, where the finite-momenta peaks are clearly established.
The observed peak positions correspond to the expected momenta close to $\hbar k=\pm (\pi/2)(\hbar/d)$~\cite{suppmat}.
In addition, Figs.~\ref{fig3}(g) and~\ref{fig3}(h), taken at $t_{\rm E}=7 \tau$ and $14 \tau$, respectively, hint at a fine structure of the emerging peaks.
This structure is a consequence of the finite TOF time $t_{\rm TOF} = 6$ms, which results in the TOF distributions being a convolution of real-space and momentum-space densities.
We sketch this situation in Fig.~\ref{fig3}(a). 
As discussed before, the peaks in $n(k)$ close to $\hbar k=-(\pi/2)(\hbar/d)$ and $+(\pi/2)(\hbar/d)$ originate from the left- and right-moving portions of the cloud, respectively.
Due to the finite $t_{\rm TOF}$, the higher-order peak of the left-moving cloud with momentum $(-\pi/2 + 2\pi)(\hbar/d)$ and the main peak of the right-moving cloud with momentum $(\pi/2)(\hbar/d)$ (and vice versa) may overlap in the TOF data.
A perfect overlap gives rise to single sharp peaks such as the ones present in the data shown in Fig.~\ref{fig3}(i), while shorter expansion times, as shown in Figs.~\ref{fig3}(g)-\ref{fig3}(h), result in a partial overlap and additional structure (see \cite{suppmat} for details).

{\it Comparison with exact time evolution.} 
We numerically model the dynamics of 1D HCBs for realistic conditions:
{\it (i)} The experimental set-up consists of many isolated 1D chains, which are not equivalent due to the 3D harmonic confinement.
Experimentally we can only measure an ensemble average over all tubes.
{\it (ii)} Both the finite temperature of the original 3D Bose-Einstein condensate as well as nonadiabaticities during the lattice loading result in a finite entropy, and thereby holes, in the initial state.
We therefore average the results over different initial product states drawn from a thermal ensemble of a harmonically trapped 3D  gas of HCBs in the atomic limit~\cite{suppmat}.
Chemical potential and temperature were calibrated to reproduce the experimental atom number and an average entropy per particle of $1.2\,k_B$~\cite{Trotzky2009}, thereby leaving no free parameters for the simulations.
To test the consistency of the approach, we compare the average density $n_j$ during the expansion with the {\it in-situ} images in Figs.~\ref{fig3}(b)-\ref{fig3}(e) and find a good agreement.
In addition, the time evolution of the half width at half maximum of the density distribution, shown in Fig.~S4 
of~\cite{suppmat}, is consistent with the ballistic dynamics as previously measured in the same experimental set-up~\cite{ronzheimer13}.

Since the momentum distribution is experimentally measured at a finite $t_{\rm TOF}$, we explicitly calculate the TOF density distributions without employing the far-field approximation~\cite{gerbier08,suppmat} and compare the results to the experimental data in Figs.~\ref{fig3}(f)-\ref{fig3}(i).
Remarkably, the positions and the structure of the peaks agree very well between experiment and theory, thereby supporting our two main results:
The central peaks indeed correspond to a large occupation of quasimomenta close to $\hbar q=\pm (\pi/2)(\hbar/d)$, i.e., to a bunching of particles around the fastest group velocities in the middle of the single-particle spectrum.
In addition, the fine structure visible for intermediate expansion times [cf.~Figs.~\ref{fig3}(g)-\ref{fig3}(h)], which becomes more apparent in the experiment when comparing different $t_{\rm TOF}$ (cf. Fig.~S2 
in~\cite{suppmat}), directly confirms the independent emergence of coherence in the left- and right-moving portions of the cloud.
Compared to the ideal case,  the presence of holes in the initial state causes a reduced  visibility of the TOF density distributions~\cite{suppmat}.
Moreover, the finite initial entropy gives rise to a crossover of one-particle correlations from a power-law decay at short distances to a more rapid decay  at long distances~\cite{suppmat}, similar to the effect of a finite temperature~\cite{rigol05b} in equilibrium.

We attribute the discrepancies between experimental and numerical results at short times, see Fig.~\ref{fig3}(f), to the weak residual $k=0$ coherence in the initial state.
Additional discrepancies may arise because of a small admixture ($\lesssim 5\%$) of doublons in the initial state~\cite{ronzheimer13,Xia2015} as well as  small residual potentials, yet we conclude that these play a minor role~\cite{suppmat}.
Compared to the previously studied time dependence of density distributions and expansion velocities~\cite{ronzheimer13}, the momentum distribution is more sensitive to such imperfections~\cite{suppmat}.
Performing a similar experiment with a single 1D system would allow the predicted scaling of $t_{\rm E}^*$ and the maximum peak height with atom number~\cite{rigol04} to be experimentally tested.

{\it  Conclusions and outlook.}
We have reported experimental evidence for a far-from-equilibrium quasicondensation at finite momenta  of expanding 1D HCBs in an optical lattice.
The expanding particles bunch at quasimomenta close to $\pm (\pi/2)(\hbar/d)$ and the analysis of TOF distributions demonstrates the existence of two independent sources of coherence.

Whether such dynamical condensation persists in higher dimensional systems constitutes an open problem, given that the existing  theoretical results are based on exact diagonalization of small systems~\cite{hen10} or time-dependent Gutzwiller simulations~\cite{jreissaty11}.
Both future experiments or advanced numerical methods (see, e.g.,~\cite{carleo12,zaletel15}) could help clarify this question.
More generally, our results raise the question of whether this type of spontaneously emerging coherence is limited to integrable systems and whether genuinely far-from-equilibrium order can also occur in generic closed many-body systems.

{\it Acknowledgments.}
We thank  F. Essler, A. Mitra and M. Rigol for helpful discussions. 
We acknowledge support from the Deutsche Forschungsgemeinschaft (DFG) through FOR 801, and the EU (AQuS, UQUAM). 
This work was supported in part by National Science Foundation Grant No.~PHYS-1066293 and the hospitality of the Aspen Center for Physics.
L.V. was supported by the Alexander-von-Humboldt foundation.

\bibliographystyle{biblev1}
\bibliography{references_arXiv}


\newpage
\phantom{a}
\newpage
\setcounter{figure}{0}
\setcounter{equation}{0}

\renewcommand{\thetable}{S\arabic{table}}
\renewcommand{\thefigure}{S\arabic{figure}}
\renewcommand{\theequation}{S\arabic{equation}}

\renewcommand{\thesection}{S\arabic{section}}


\begin{center}
\Large
\hypertarget{pagesupp}{Supplemental material}
\end{center}

\label{pagesupp}

\section{Experimental details}

\subsection{Increasing the visibility of higher order peaks}
\label{sec:flash}
\begin{figure}[!bt]
\begin{center}
\includegraphics[width=\columnwidth]{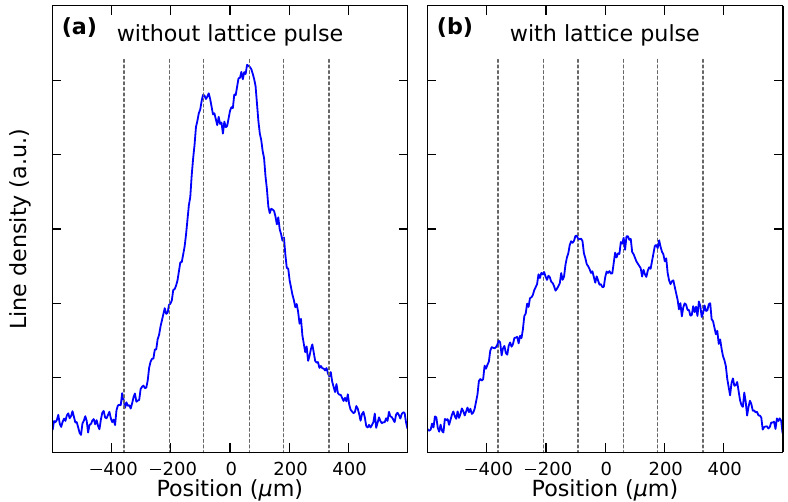}
\caption{ {\it Experimental TOF density distributions.}  
Comparison between a measurement (a) without additional lattice pulse and (b) with additional pulse before the TOF sequence (in both cases $t_{\rm E}=22\,\tau$ and $t_{\mathrm{TOF}}=12\,\mathrm{ms}$).
Vertical lines are guides to the eye.
Density distributions are averaged over 10 experimental realizations.
}
\label{fig_LatticeFlashing}
\end{center}
\end{figure}

By switching off the lattice instantaneously at the beginning of the time-of-flight (TOF) sequence, each quasimomentum $\hbar q$ is projected onto a superposition of states with free-space momentum $\hbar k$ (with $k=q$) and higher-order (Bragg) contributions with momenta $\hbar (k \pm n 2\pi / d)$, $n \in \mathbb{N}$.
In order to increase the visibility of the higher-order peaks, we increase the intensity of the $x$-lattice for $5\,\mu s$ to approximately  $33E_r$ immediately before releasing the atoms from the lattice.
This time is long enough to lead to a narrowing of the on-site (Wannier) wavefunctions, resulting in a broadening of their momentum distributions, while being short enough to have no significant influence on the quasimomentum distribution. This leads to a broadening of the envelope of the momentum distribution observed after time of flight and thereby enhances the visibility of the higher order (Bragg) peaks, as shown in Fig.~\ref{fig_LatticeFlashing}.

\subsection{Analysis of observed peak positions during time of flight}

The expected signatures of the quasicondensation at finite quasimomenta close to $\pm (\pi/2)(\hbar/d)$ are distinct peaks in the density distribution after the TOF sequence.
Since the quasicondensates form independently in the left- and right-moving portions of the cloud during the sudden expansion, the peaks observed in the TOF distribution can be interpreted as the sum of two independent contributions originating from different points in space denoted by $d_{\rm L}^0$ and $d_{\rm R}^0$, respectively.
For specific TOF times, different momenta stemming from different positions in the cloud can become superimposed and appear as one larger peak in the measured TOF density distribution.
To demonstrate this effect, we extract the peak positions for various different TOF times $t_{\mathrm{TOF}}$ with a simple fit function consisting of six Gaussian peaks on top of a broad background.

The six small Gaussian peaks are grouped into two groups corresponding to the left- and right-moving portions of the cloud.
Each group contains one peak at position $d_{\rm L/R}$ that corresponds to $d_{\rm L/R}^0+t_{\mathrm{TOF}} \cdot \hbar k_{\rm L/R}/m$, respectively, and the additional higher order peaks at positions $d_{\rm L/R}\pm d_{2\pi}$, where $d_{2\pi} = t_{\mathrm{TOF}} \cdot 2\pi \hbar/(d m) $.
In order to further reduce the number of free fitting parameters, we assume all six Gaussian peaks to have the same width $w_G$, but allow for individual amplitudes $A_{\rm L/R}^i$.
The complete fit function $F(x,\dots)$ is given by 
\begin{equation}
\begin{split}
F(x,\dots)= \; & P_{b}(x,A_{b},d_{b},w_{b})\\
 +\,\, &P_{\rm L}(x,A_{\rm L}^{-1},A_{\rm L}^{0},A_{\rm L}^{1},d_{\rm L},w_G)\\
 +\,\, &P_{\rm R}(x,A_{\rm R}^{-1},A_{\rm R}^{0},A_{\rm R}^{1},d_{\rm R},w_G)
\end{split}
\end{equation}
with the broad background peak 
\begin{equation}
P_{b}(x,\dots) = A_{b} e^{\left(-\frac{(x-d_{b})^2}{2 w_{b}^2}\right)}
\end{equation} and the groups of small peaks 
\begin{equation}
\begin{split}
P_{\rm L/R}(x,\dots) = \, &A_{\rm L/R}^{-1}  e^{\left(-\frac{(x-d_{\rm L/R}-d_{2\pi})^2}{2 w_G^2}\right)}\\
+\, &A_{\rm L/R}^{0}  e^{\left(-\frac{(x-d_{\rm L/R})^2}{2 w_G^2}\right)}\\
+\, &A_{\rm L/R}^{1}  e^{\left(-\frac{(x-d_{\rm L/R}+d_{2\pi})^2}{2 w_G^2}\right)}.
\end{split}\,
\end{equation}

Due to the additional sudden switching off of the magnetic field used to address the Feshbach resonance and the resulting eddy currents, there are additional position-dependent forces acting on the atoms during the initial stage of the TOF sequence.
In order to correctly map the observed distances after time of flight onto the initial momenta, we apply a correction that is determined by monitoring the evolution of the TOF peaks of an equilibrium $q=0$ Bose-Einstein condensate released from the lattice, where the momentum composition is known to consist of peaks at $\pm n 2\pi (\hbar /d)$ with $n\in \mathbb{N}$.
    
\begin{figure}[!bt]
\begin{center}
\includegraphics[width=\columnwidth]{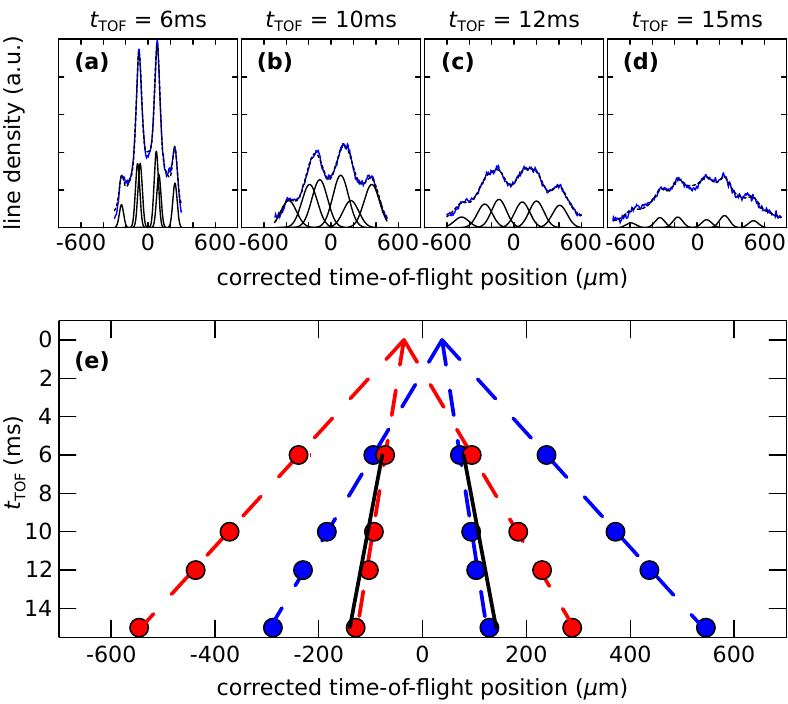}
\caption{
{\it Peak positions as a function of time-of-flight time}.
(a)-(d) Density distributions obtained for a fixed expansion time of $t_{\rm E} = 36\tau$ (blue solid lines). 
Black dashed lines are fits with a sum of six Gaussian peaks and a broad background, black solid lines show only the contributions of the six peaks.
All measured density distributions are averaged over 9 to 10 experimental realizations.
(e) Extracted peak positions for $t_{\rm E} = 36\tau$ (dots).
Dashed lines are linear fits, solid black lines show the expected slope for peaks traveling with momentum $\pm (\pi/2)(\hbar/d) $.
Error bars in the position are smaller than the data points.
}
\label{fig_PeakPos}
\end{center}
\end{figure}

Figures~\ref{fig_PeakPos}(a)-\ref{fig_PeakPos}(d) show the density distributions after an expansion time $t_{\rm E} = 36\tau$ in the lattice and varying TOF times $t_{\rm TOF}$ together with the resulting fits.
Even though the fits result in strongly varying peak amplitudes, they nonetheless serve to faithfully identify the peak positions.

In Fig.~\ref{fig_PeakPos}(e), we show the extracted positions of the six peaks (red and blue circles) for a fixed expansion time $t_{\rm E} = 36\tau$.
The red circles are associated with the component at negative momentum $\hbar k_{\rm L}$ and its corresponding higher order peaks at $\hbar (k_{\rm L}\pm 2\pi /d)$, the blue circles are associated with the positive momentum $\hbar k_{\rm R}$ and higher order peaks.

The dashed lines are linear fits, their slopes represent the extracted velocities. For comparison, the solid black lines show the theoretically expected slope for momenta $\pm (\pi/2)(\hbar /d)$. 
The peaks do indeed propagate with momenta close to the ideal expectation; the small shift towards smaller momenta 
can be attributed   to the finite interaction strength $U/J<\infty$ used in the experiment (see Sec.~\ref{sec:dev}).

The extrapolation to $t_{\mathrm{TOF}} = 0$ shows that the two momentum groups indeed originate from two different points in space in the initial cloud, with the left-moving part starting further to the left and vice versa. 

At $t_{\mathrm{TOF}} = 6\,\mathrm{ms}$ in Fig.~\ref{fig_PeakPos}(e), we can also observe how multiple peaks can come to overlap during the TOF evolution, resulting in the observation of fewer but stronger peaks, explaining the number of peaks observed in, e.g., Fig.~\ref{fig_PeakPos}(a).


\section{Time-of-flight distribution without the far-field approximation}

After the lattice flashing, each site of the lattice can be described by the Wannier function
$w_0(r) $, which we approximate by a Gaussian of width $\sigma$:
\begin{equation}
w_0(r) = \frac{1}{\sqrt{\sigma \sqrt{\pi}}} e^{-r^2/(2\sigma^2)}.
\end{equation}
As a consequence of the lattice flashing method described in Sec.~\ref{sec:flash}, $\sigma \ll d$.
After switching off the lattice, each Wannier function freely evolves in time.
We calculate the free time evolution in momentum space,
${\rm FT}[w_0(r,t_{\rm TOF})] = {\rm FT}[w_0(r)] e^{-i E_k t_{\rm TOF}/\hbar}$,
where FT denotes the Fourier transform and $E_k = (\hbar k)^2 /(2m)$ is the free-space single-particle dispersion.
For the propagation in real space, we get
\begin{equation}
w_0(r,t_{\rm TOF}) = \frac{1}{\sqrt{ ( \sigma+ i \frac{\hbar t_{\rm TOF}}{m \sigma } )  \sqrt{\pi} }} e^{-\frac{r^2}{2\sigma^2 + i \frac{2\hbar t_{\rm TOF}}{m}}}.
\end{equation}
The total density profile $n_{\rm TOF}(r) \equiv n_{\rm TOF}(r,t_{\rm TOF}) =  \langle  \hat \psi^\dagger(r) \hat  \psi(r) \rangle$ at a given $t_{\rm TOF}$ is a sum over contributions from all (occupied) lattice sites,
\begin{equation}
n_{\rm TOF}(r) = \sum_{r_j, r_l} w_0^*(r-r_j,t_{\rm TOF}) w_0(r-r_l,t_{\rm TOF})
\langle  \hat a_{r_j}^\dagger \hat  a_{r_l} \rangle,
\end{equation}
where the field operator $\hat \psi(r)$ was expanded in time-evolved Wannier functions.
In the case of one-dimensional (1D) lattice dynamics studied here, off-diagonal correlators $\langle \hat a_{r_j}^\dagger \hat a_{r_l} \rangle$ are nonzero along a single direction only, hence we define $x \equiv r/d$ and $\mu \equiv r_\mu$.
The density  can be written in a compact form by introducing two dimensionless parameters
$\alpha = 1/[(\sigma/d)^2 + (\hbar t_{\rm TOF} / (m\sigma d))^{2}]$ and $\beta = \alpha \hbar t_{\rm TOF} /(m \sigma^2)$, resulting in 
\begin{eqnarray}
n_{\rm TOF}(x) &=& \frac{1}{d} \sqrt{\frac{\alpha}{\pi}} e^{-x^2 \alpha} \sum_{j,l} e^{x(l+j) \alpha} e^{-\frac{1}{2} (l^2+j^2) \alpha} \times \nonumber \\
& & \times \; \;  e^{-i x (l-j) \beta} e^{i(l^2-j^2) \frac{\beta}{2}}
\langle \hat a_j^\dagger \hat a_l \rangle.
\end{eqnarray}
This generalizes the result of Gerbier {\it et al.}~\cite{gerbier08}, where only the phase factors have been taken into account.
In our experiment $\sigma = 50$ nm, i.e., $\sigma/d \approx 0.13$.
The central task is therefore to calculate the correlators $\langle \hat a_j^\dagger \hat a_l \rangle$, for which we
provide details in the following Secs.~\ref{secLnum} and \ref{sec:corr}.


\section{Details of the numerical simulations}

\subsection{Realistic modeling of the initial state}
\label{secLnum}

We model the experimental set-up by calculating the exact time evolution of many 1D systems of hard-core bosons (HCBs)~\cite{rigol04,vidmar13} that differ by particle numbers $N$ and their distribution in the initial state.
The generic initial state within a single chain is a product state
\begin{equation} \label{i_state}
|\varphi_0\rangle = \prod_{j \in {\cal J}} \hat a^{\dagger}_j |\emptyset\rangle,
\end{equation}
where $\cal J$ does not necessarily represent a sequence of neighboring sites.
The  initial states are drawn from a thermal distribution of HCBs in the atomic limit ($J=0$) of a 3D cloud and  the  final result of the simulations is
obtained by summing up the data for  initial states with different $\cal J$.

The probability to find a particle at a given lattice site is given by the following partition function,
\begin{equation} \label{partition}
Z = \sum_{n=0,1} e^{-\beta (n V(x,y,z) - n\mu_0)},
\end{equation}
where $\mu_0$ represents the chemical potential, $\beta=1/(k_{\rm B}T)$ is the inverse temperature and the harmonic confinement is given by
\begin{equation}
V(x,y,z)= \frac{m\omega_{x,y}^2}{2} \left[x^2+y^2+(\Lambda z)^2\right],
\end{equation}
with $\omega_{x,y}\equiv \omega = 2 \pi \times 60$~Hz and the aspect ratio $\Lambda=\omega_z/\omega_{x,y}=2.63$.
The 1D systems are aligned along the $x$-direction and the results are integrated over chains at different positions $(y,z)$.
Due to the harmonic confinement, chains at different transverse positions can have vastly different atom numbers in the range $N\in [0,80]$.
Clearly, for $T \to 0$, the initial state in the 3D cloud consists of an ellipsoid of singly-occupied sites without any holes.
A finite-entropy initial state, on the other hand, contains a finite density of holes that increases when moving outwards.
Equation~(\ref{partition}) leads to the global density 
\begin{equation}
n(x,y,z) = \frac{1}{e^{\beta(V(x,y,z)-\mu_0)}+1}.
\end{equation}

\begin{figure}[!]
\begin{center}
\includegraphics[width=0.90\columnwidth]{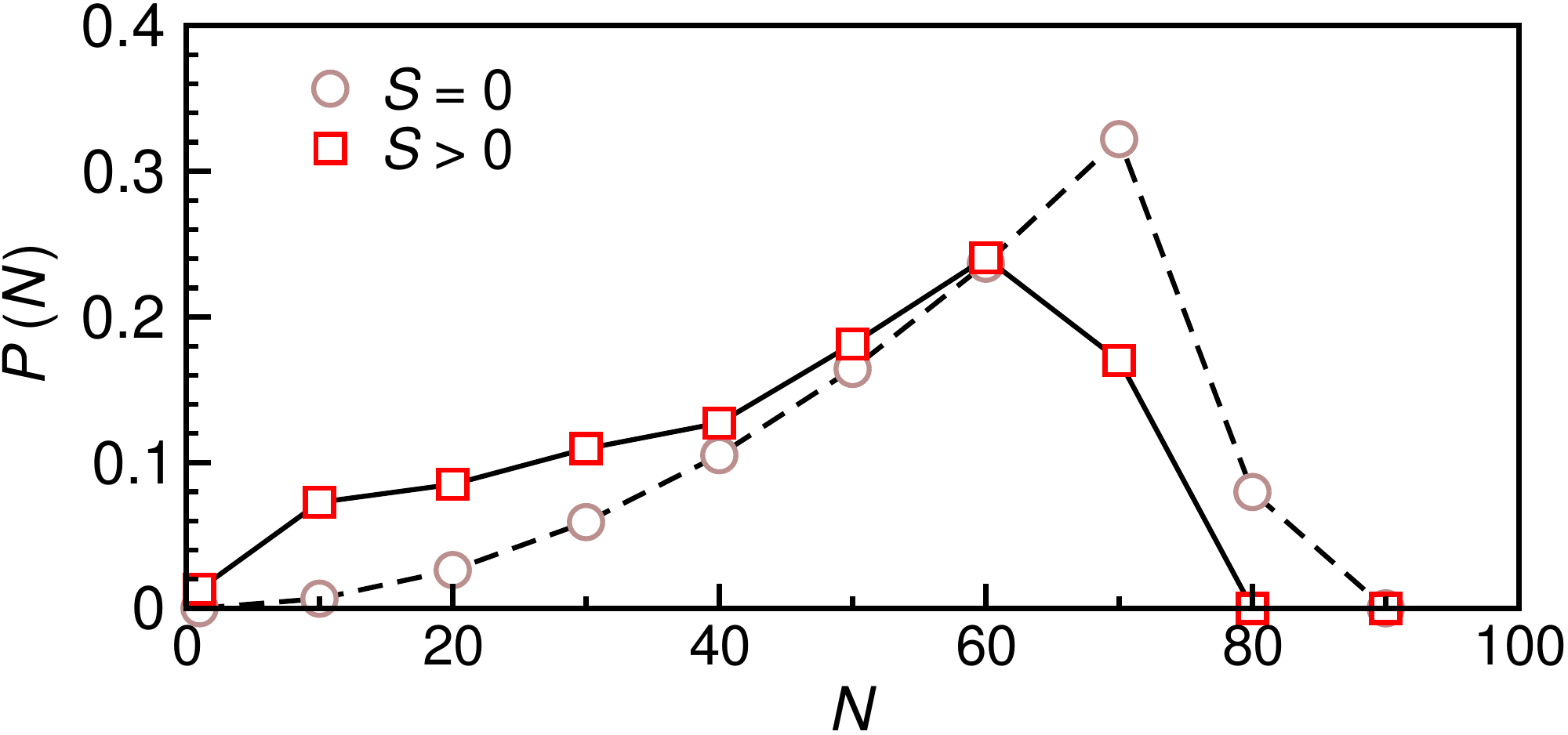}
\caption{
{\it Distribution of probabilities $P(N)$  to find particles in chains with particle number $N$.}
We compare the zero-entropy ($S=0$) and the finite-entropy case ($S/N_{\rm 3D} =1.2 \; k_B$).
We use a binning with $\Delta N = 10$.
} 
\label{fig_ParticleDistribution}
\end{center}
\end{figure}

We determine $\beta$ and $\mu_0$ by matching the total particle number $N_{\rm 3D}$ of the 3D cloud and the average entropy per particle $S/N_{\rm 3D}$ to  experimental conditions.
The average entropy per particle is defined as
\begin{equation}
S/N_{\rm 3D} = \sum_{i=(x,y,z)} S_i/N_{\rm 3D},
\end{equation}
where
\begin{equation}
S_i = -k_B \left( n_i \log{n_i} + (1-n_i) \log{(1-n_i)} \right)
\end{equation}
and $n_i$ represents the probability to find a particle at site $i$.
In our experiment, we have $N_{\rm 3D} = \sum_{x,y,z} n(x,y,z) \approx 0.9 \times 10^5$ and we estimate $S/N_{\rm 3D}\approx 1.2 \; k_B$~\cite{Trotzky2009}.
A typical configuration $\cal J$ for $N=50$ is shown in Fig.~\ref{fig_corr}(a2).
Note that we also symmetrize each configuration by taking its  mirror image (with respect to the central site of the originally occupied region).

In Fig.~\ref{fig_ParticleDistribution} we plot the distribution of probabilities $P(N)$ to find particles in chains with a given number of particles $N$.
This information reveals the relative contribution of chains with a given particle number to the dynamics.
The distribution for  $S=0$ is shown with circles.
In this case, the 3D state corresponds to a perfectly filled ellipsoid $R_0^2 = x^2 + y^2 + (\lambda z)^2$ with $R_0/d=38$ and, as a consequence,
$P(N) \propto N^2$ for $N \lesssim 70$.
As a general trend, the distribution shifts towards smaller $N$ with increasing entropy (squares in Fig.~\ref{fig_ParticleDistribution}).

In our calculations, we simulate the dynamics of representative chains with particle numbers $N \in \{ 1,10,20,30,40,50,60,70\}$ and the relative weights shown in Fig.~\ref{fig_ParticleDistribution} for $S/N_{\rm 3D}  = 1.2 \; k_B$.
All together we average the data over 100 different initial configurations and we have verified that the sampling over this subset of all possible $N$ is sufficient.
In all data for {\it in-situ} and TOF density distributions shown in the figures, we convolve the numerical data with a Gaussian filter of width $7d$ to account for the finite experimental imaging resolution.

\begin{figure}[!t]
\begin{center}
\includegraphics[width=0.90\columnwidth]{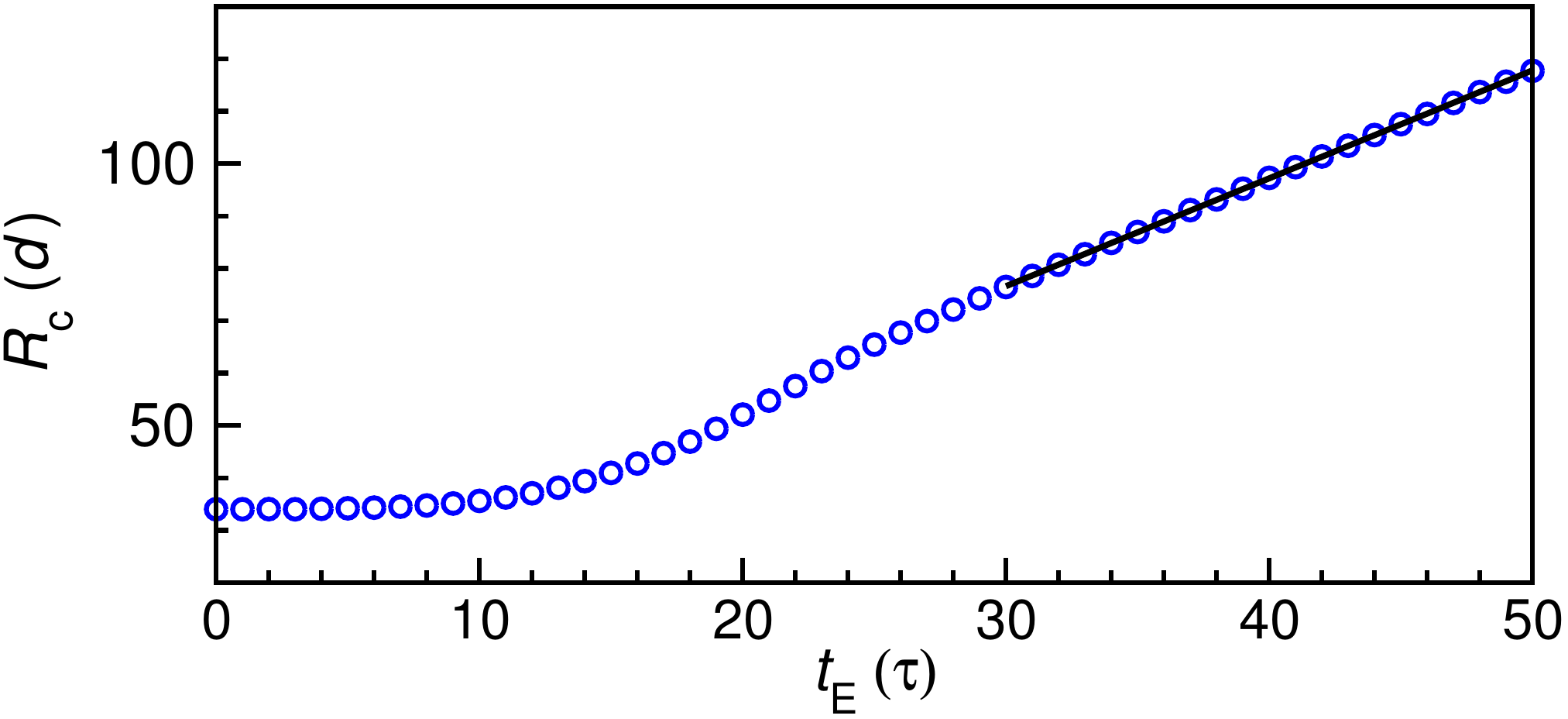}
\caption{
{\it Time dependence of the core radius $R_{\rm c}(t_{\rm E})$ at $S/N_{\rm 3D}  = 1.2 \; k_B$.}
The solid line represents the fit function $R_c(t_E)= v_c t_{\rm E} +\gamma$ with $v_c= 2.06(d/\tau)$ and $\gamma=14.8$,  obtained by fitting to the numerical data in the time interval $t_{\rm E}/\tau \in [30,50]$.
} 
\label{fig_Rcore}
\end{center}
\end{figure}

\begin{figure*}[!]
\begin{center}
\includegraphics[width=1.99\columnwidth]{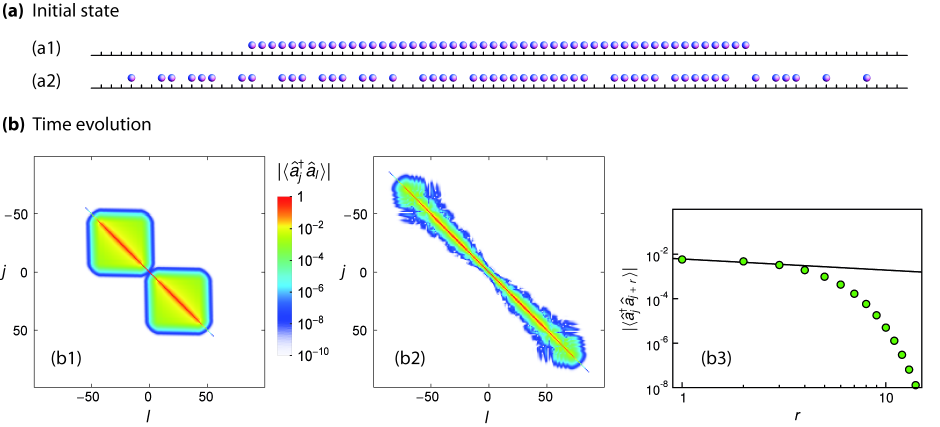}
\caption{
{\it Influence of holes in the initial state on the one-particle correlations.}
(a) Snapshots of  atom configurations in the initial state ($N=50$ in all cases) for (a1) an ideal initial state $|\psi_0\rangle$, Eq.~(\ref{idealstate}), and for (a2) a typical initial state in Eq.~(\ref{i_state}), generated at $S/N_{\rm 3D} =1.2 \; k_B$.
(b) Modulus of one-particle density matrix $|\langle \hat a^\dagger_j \hat a_{l}\rangle|$ for (b1) $t_{\rm E}=10 \tau$, emerging from the ideal initial state in (a1), and for (b2) $t_{\rm E}=14 \tau$, emerging from the symmetrized initial state consisting of the state in (a2) and its mirror image, obtained by
reflection with respect to the central site. 
(b3) Spatial decay of $|\langle \hat a^\dagger_j \hat a_{j+r}\rangle|$ at $j=68$ of the (non-symmetrized) state shown in (a2) at $t_{\rm E}=14 \tau$.
The solid line is the function $f(r) = \alpha/r^{1/2}$ (with $\alpha=6.1 \times 10^{-3}$) to illustrate that the decay at large distances is faster than a power law.
} 
\label{fig_corr}
\end{center}
\end{figure*}

Figure~\ref{fig_Rcore} shows that the core radius (i.e., the half width at half maximum) increases linearly in time according to $R_{\rm c}(t_{\rm E}) = v_{\rm c}t_{\rm E}+ {\rm \it const}$
with the core velocity very close to $v_{\rm c}=2 (d/\tau)$, in agreement with~\cite{ronzheimer13}.
This suggests that the presence of  holes in the initial state does not affect the ballistic expansion of 1D HCBs.

\subsection{Correlation functions from the exact time evolution of a 3D cloud of HCBs}
\label{sec:corr}

We proceed with the analysis of the one-particle correlations $\langle \hat a^\dagger_j \hat a_{l}\rangle$ during the time evolution.
We compare a  representative chain at $S/N_{\rm 3D}=1.2 \; k_B$ to the ideal initial state, both with $N=50$ particles.
The ideal initial state is (see the main text)
\begin{equation} \label{idealstate}
|\psi_0\rangle = \prod_{j \in L_0} \hat a^{\dagger}_j |\emptyset\rangle,
\end{equation}
where the length of the region $L_0$ equals $N$. 
The distribution of particles in $|\psi_0\rangle$ is displayed in Fig.~\ref{fig_corr}(a1), while the particle configuration of a typical chain at $S/N_{\rm 3D}=1.2 \; k_B$ is shown in Fig.~\ref{fig_corr}(a2).

The modulus of the one-particle density matrix $|\langle \hat a^\dagger_j \hat a_{l}\rangle|$ resulting from $|\psi_0\rangle$ is shown in Fig.~\ref{fig_corr}(b1) for $t_{\rm E}=10 \tau$.
The two squares visualize the two sources of coherence emerging in the left- and right-moving cloud~\cite{rigol04}, corresponding to the peaks in $n(q)$ at $\hbar q=-(\pi/2)(\hbar/d)$ and $\hbar q= (\pi/2)(\hbar/d)$ in Fig.~\ref{fig2}(c), 
respectively.
Within the squares, the correlations decay according to a power law (cf.~Fig.~\ref{fig2}(d)
in the main text).

The modulus of the one-particle density matrix $|\langle \hat a^\dagger_j \hat a_{l}\rangle|$ for the  chain displayed in Fig.~\ref{fig_corr}(a2)  is shown in Fig.~\ref{fig_corr}(b2) at $t_{\rm E}=14\tau$.
As this example illustrates, the correlations in chains with holes  show a less regular pattern, compared to the ideal situation.
However, the calculated phase differences between neighboring sites exactly equal $\pm \pi/2$, identical to the ideal case shown in the inset of Fig.~\ref{fig2}(d)
of the Letter.
In Fig.~\ref{fig_corr}(b3) we plot $|\langle \hat a^\dagger_j \hat a_{j+r}\rangle|$ as a function of $r$ at $j=68$ and the comparison to the function $f(r)=\alpha/r^{1/2}$ clearly shows that, taking holes in the initial state into account, the correlations decay much faster than in the ideal case for large distances.
Therefore, even though signatures of the finite-momentum quasicondensation can still be detected in a system with a  finite initial entropy, the correlations have become short ranged.
This is equivalent to the behavior of equilibrium 1D Bose gases at any finite temperature~\cite{giamarchibook}, where temperature cuts off the ground-state power-law decay.
The temperature, or the amount of entropy, thereby sets the crossover scale from power-law to exponentially decaying correlations.

\subsection{Other sources for a loss of coherence}
\label{sec:dev}
While the overall agreement between the experimental data and our numerical simulations is quite good (see Figs.~\ref{fig3}(f)-\ref{fig3}(i) 
of the Letter),
there are nevertheless small discrepancies.
For instance, the  experimental data exhibits broader peaks and a lower visibility even at the longest expansion times [see Fig.~3(g) in the main text]), which arise from additional sources of decoherence present in the experiment. 
We have numerically investigated the effect of the following deviations from ideal conditions:
(i) the finite interaction strength $U/J=20<\infty$ used in the experiment,
(ii) the presence of small residual potentials,
(iii) the presence of a small admixture of doublons in the initial state, and
(iv) the influence of coherence in the initial state, i.e., the presence of a maximum in $n(q)$ at $q=0$.
Away from the HCB limit [i.e., investigating the effects (i) and (iii)], one needs to resort to time-dependent density matrix renormalization group (tDMRG) simulations \cite{vidal04,daley04,white04}.
For the tDMRG method, a realistic modeling with particle numbers as large as in the experiment and sufficiently long times is currently unfeasible.

\begin{figure}[!]
\begin{center}
\includegraphics[width=0.90\columnwidth]{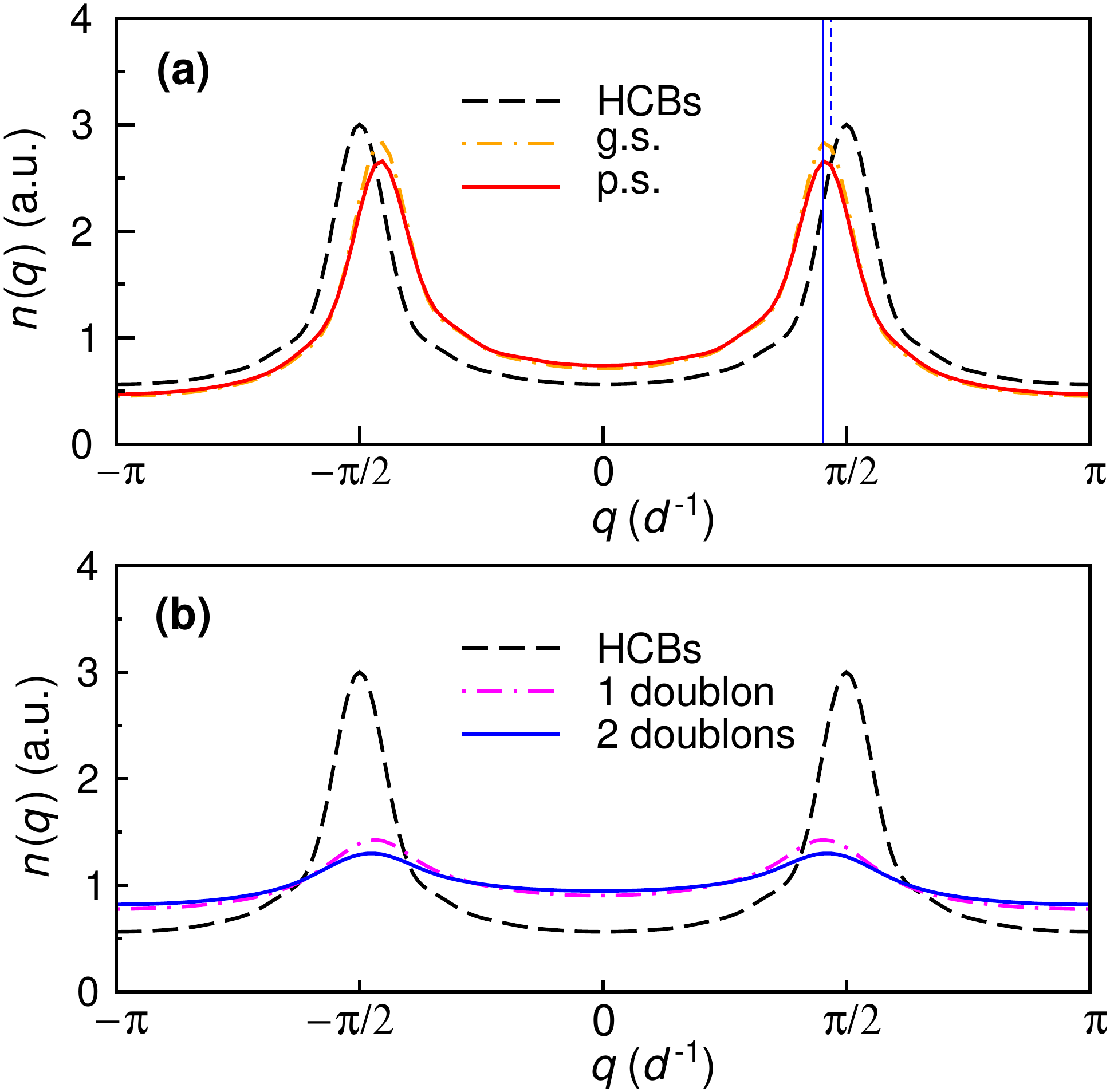}
\caption{
{\it Quasimomentum distribution of HCBs and interacting bosons at $U/J=20$.}
We compare the dynamics at $t_{\rm E}=(N/4)\tau$ and $N=16$.
Results for  HCBs are shown with dashed lines in both panels. In (a), we show $n(q)$ for the expansion from 
the  ground state (g.s.)~at $U/J=20$ and the initial product state (p.s.)~$|\psi_0\rangle$ from Eq.~(\ref{idealstate}).
The solid and dashed vertical lines denote the estimate $q'=1.42/d$ and $1.47/d$, respectively, for the quasimomenta at the shifted maxima.
In (b) we show $n(q)$ for the expansion from an initial product state with a single doublon and hole (dash-dotted line; averaged over 184 configurations) and with two doublons and two holes (solid line; averaged over 552 configurations).
}
\label{fig_finiteU}
\end{center}
\end{figure}

The dynamics of the quasimomentum distribution in the sudden expansion at $U/J<\infty$ starting from the correlated ground state at density one has been investigated in~\cite{rodriguez06}.
It has been observed that the position of the maxima moves to smaller values of $\hbar |q| < (\pi/2)(\hbar/d) $.
As a heuristic estimate for the peak position~\cite{rigol04} and to account for the conservation of total energy in this quantum quench, 
one can assume that all particles condense at the point in the single-particle dispersion corresponding to the
average energy per particle. This results in the  condition $E/N = -2J \cos{(q d)}$.
In the hard-core case, the average energy per particle vanishes, and hence one obtains $\hbar q=\pm (\pi /2)(\hbar/d)$. 

For expansion from the correlated {\it ground state} at $U/J=20$, we express the ground-state energy $E_{\rm g.s.}=-4NJ^2/U$ by  second-order perturbation theory around the $J/U=0$ limit,  arriving at
$\hbar q'=\arccos{(2J/U)}(\hbar/d)$.
Our tDMRG results for $n(q)$ at $U/J=20$ are shown in Fig.~\ref{fig_finiteU}(a) and the  estimate for the position of the peak $\hbar q'= \arccos{(2/20)}(\hbar/d) = 1.47(\hbar/d)$ is in good agreement with the numerical data~\cite{rodriguez06}.

In our experiment, we ideally start from the {\it product state}~(\ref{idealstate}) of one boson per site.
The quasimomentum distribution for the latter initial state, shown in Fig.~\ref{fig_finiteU}(a), also exhibits a slight shift of the maxima to smaller absolute values of the peak position.
One can understand this shift from the  two-component picture discussed in Sec.~IV of Ref.~\cite{sorg14} as follows:
We separate the gas into the  ballistically expanding unbound particles and inert doublons. The 
doublons form on time scales for which the cloud has not yet considerably expanded into an empty lattice \cite{ronzheimer13}.
Upon opening the trap, the doublons undergo the quantum distillation process~\cite{hm09,muth12,bolech12,Xia2015} that makes them accumulate in the center of the lattice, such that
on the time scales probed by the experiment,  they do not contribute to the quasicondensation (see the supplemental material of \cite{ronzheimer13}).
The number of doublons $\braket{d} = 4NJ^2/[U(U-6J)]$ can be estimated from  second-order perturbation theory (see Eq.~(39) in~\cite{sorg14}).
The condition for the position of the dynamically formed quasicondensate can then be formulated as
$E'/N' = -2J \cos{(q'd)}$,
where $E'=-\braket{d}U$ and $N'=N-2\braket{d}$ are the energy and the number of ballistically expanding particles, respectively.
For $U/J=20$, we find $\hbar q' = 1.42(\hbar/d)$, in very good agreement with the numerical data shown in Fig.~\ref{fig_finiteU}(a).

The main effect of a finite $U/J=20$ is thus the slight shift of the peak positions in the dynamical quasicondensation of $n(q)$.
On the other hand, the shape of the peaks does not change considerably.
Small residual potentials also cause a time-dependent shift of the position of the maxima~\cite{mandt13}, but do not lead to a broadening of these peaks.
Our numerical results for HCBs (not shown here) are consistent with these predictions.

From our previous work~\cite{ronzheimer13}, we estimate that a small fraction of   $\lesssim 5\%$ doublons is present in the initial state.
Unfortunately, this regime of small doublon fractions can hardly be accessed with the tDMRG method as it would require large particle numbers, severely restricting the accessible time scales.
We have performed tDMRG simulations with one or two doublons in the initial state with an overall particle number of $N=16$.
This translates into a fraction of doublons $> 5\%$ and we observe that in these cases (after averaging over many initial configurations) the core expansion velocity $v_{\rm c}$ drops
below the experimentally measured value, inconsistent with the results of Ref.~\cite{ronzheimer13}.
It is nonetheless obvious from our tDMRG simulations shown in Fig.~\ref{fig_finiteU}(b) that the presence of doublons in the initial state causes a significant reduction of the visibility of the finite-momentum peaks in $n(q)$.
Thus deviations between our experimental data and the results of the numerical simulations for HCBs can be partially attributed to doublons.

Finally, we discuss the influence of the existence of some short-range coherence in the system present at $t_{\rm E}=0$ on the 
expansion dynamics and the TOF distribution.
Such $q=0$ short-range coherence is evident in the experimental data shown in Fig.~\ref{fig3}(d) of the main text,
and results in additional peaks in the TOF distribution beyond those expected from the emergent quasicondensates at finite momenta.
This initial $q=0$ coherence, which stems from an imperfect state preparation, clearly decays rather fast and one can see from two examples that it is not detrimental to the formation of
quasicondensates at finite momenta at sufficiently long expansion times.

\begin{figure}[!]
\begin{center}
\includegraphics[width=0.93\columnwidth]{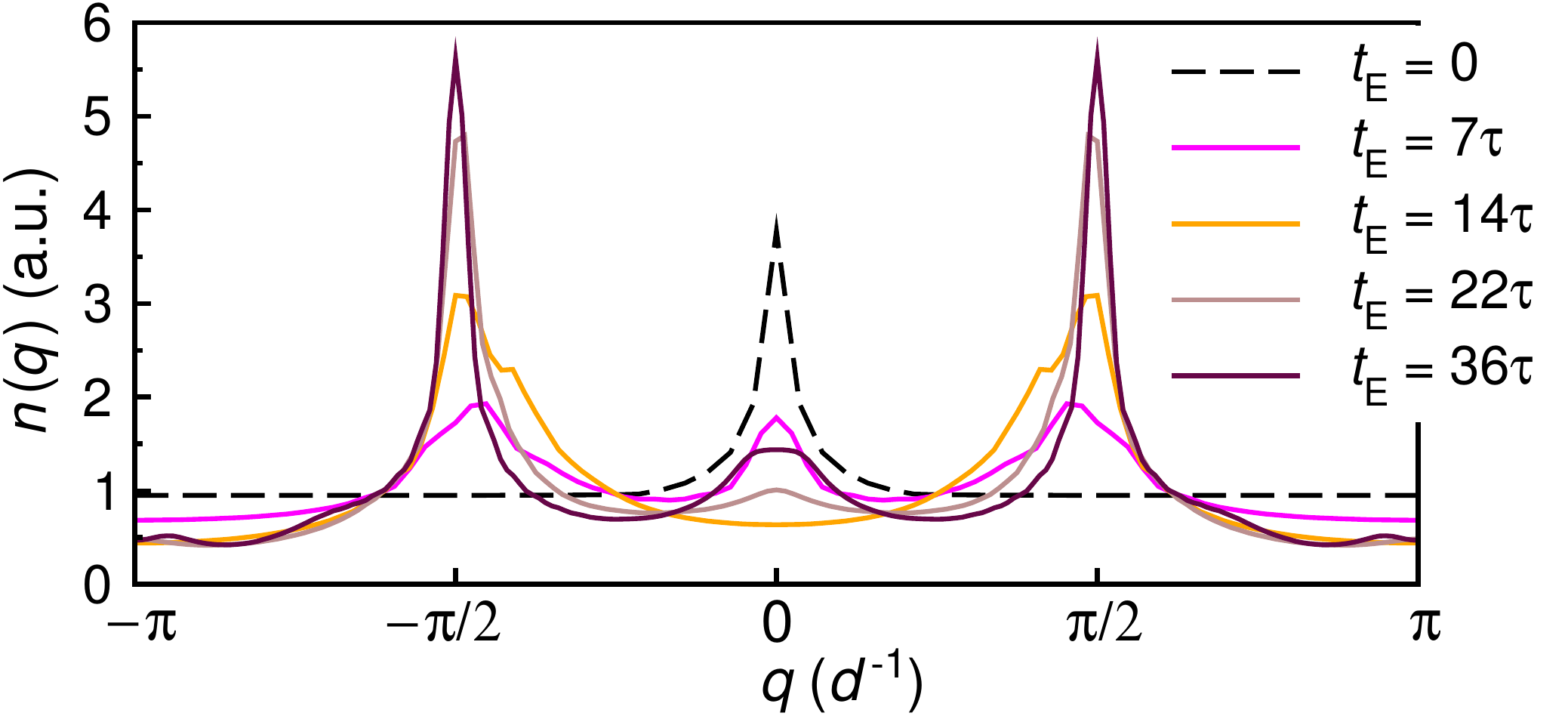}
\caption{
{\it Sudden expansion of HCBs from an initial state with an incommensurate density.}
We show $n(q)$ for the expansion from a box trap with $N=45$ particles and an initial density $n=0.9$,
plotted for expansion times $t_{\rm E}/\tau=0,7,14,22,36$.
}
\label{fig:sm-new}
\end{center}
\end{figure}

First, in the expansion from the correlated ground state of a Mott insulator at $U/J<\infty$, there is some short-range coherence present at $t_{\rm E}=0$ already.
Nonetheless, the results of Ref.~\cite{rodriguez06}, where the expansion from the correlated Mott-insultating states was studied,  clearly show that sharp peaks
at finite momenta form dynamically in $n(q)$. Second, consider the sudden expansion of hard-core bosons from a box trap,
yet with a density below unity in the initial state and from a correlated state  such as the ground state in that box.
For this case, we show the time evolution of the quasimomentum distribution $n(q)$ 
in Fig.~\ref{fig:sm-new}. As expected, there is a maximum in $n(q)$ at $q=0$ at $t_{\rm E}=0$, which quickly diminishes in favor of the quasicondensation peaks that emerge at $\hbar q=\pm \pi/2 (\hbar/d)$.
The quasicondensation peaks ultimately sit at the same position as if the expansion had started from the state given in Eq.~\eqref{idealstate}, but the distribution is not symmetric around $\hbar q=\pm \pi/2 (\hbar/d)$.
Note that the reappearance of the peak at $q=0$ observed at the longest expansion time $t=36\tau$ is a precursor to the dynamical fermionization occurring at very long expansion times~\cite{rigol05,minguzzi05}.

\subsection{Current-carrying states and power-laws}

Both our expanding bosons and the melting domain-walls realize current-carrying states: In the first case, we are dealing with particle 
currents, in the latter case, with spin currents.
Based on the scaling solution for the domain-wall melting \cite{antal99}, one can associate a current $j(\tilde x)$ and a density $n(\tilde x)$ to each point in
space where $\tilde x$ is the rescaled coordinate (see the main text). In the limit of large systems, each region described by the rescaled 
coordinate corresponds to an extended region in terms of the original spatial coordinate $x$, suggesting that one can use the local
density approximation to describe the system. In this picture, we seek  homogeneous reference systems  that have the same density $n=n(\tilde x)$ 
and the same current $j=j(\tilde x)$ as the expanding cloud at the rescaled position $\tilde x$.
We then need two parameters to match these conditions. If we assume that the reference systems have periodic boundary conditions, then
these two parameters are the chemical potential $\mu$, fixing the density,  and the flux $\phi$ through the ring, fixing the current.
By numerical comparison we find that the chemical potential is linear in $x$ while the flux $\phi$ is independent of $\tilde x$ and always
given by $\pi/2$ (details will be published elsewhere).
 
The question of power-law correlations can thus  be reformulated in two ways: First, do the (homogeneous) reference systems on which $\phi$
enforces a nonzero current have power-law correlations and second, does this also apply to inhomogeneous systems. For simple
1D spin systems that allow a mapping onto free fermions and that have conserved currents, the answer to both questions is yes, according to the 
analysis of \cite{antal98,antal99}. 
The extension to interacting systems with or without exactly conserved currents or including integrability breaking terms will be addressed in a
future publication.
Examples for such systems are the spin-1/2 XXZ model or the 1D Bose-Hubbard model. 

Beyond specific spin-1/2 chain models, related questions pertaining to the existence of power-law correlations in excited states that carry a finite
current have been addressed in various contexts, including highly excited states in  integrable 1D systems \cite{pantil13,fokkema14}, quantum magnetism \cite{antal98,antal97,gobert05}, hydrodynamics~\cite{schmittmann95} and statistical physics \cite{spohn83,katz83,schmittmann95}. Thus our work connects a broad range of branches of theoretical physics.

\newpage

\end{document}